\documentclass[aps]{revtex4}

\usepackage{graphicx}% Include figure files
\usepackage{listings}

\begin{document}

\title{Python Unleashed on Systems Biology}
\author{Christopher~R.~Myers$^1$, Ryan~N.~Gutenkunst$^2$, and James.~P.~Sethna$^2$}
\affiliation{$^{1}$Cornell Theory Center, Cornell University, Ithaca, NY 14853, USA\\
$^{2}$Laboratory of Atomic and Solid State Physics, Cornell University, Ithaca, NY 14853, USA}

\date{\today}

\begin{abstract}
We have built an open-source software system for the modeling of 
biomolecular reaction networks, \verb+SloppyCell+, which is written in 
Python and makes substantial use of third-party libraries for numerics, 
visualization, and parallel programming.  We highlight here some of 
the powerful features that Python provides that enable \verb+SloppyCell+ 
to do dynamic code synthesis, symbolic manipulation, and parallel 
exploration of complex parameter spaces.
\end{abstract}

\maketitle

\section*{Introduction}
\label{Sec-Introduction}

A central component of the emerging field of systems biology is the
modeling and simulation of complex biomolecular networks, describing
the dynamics of regulatory, signaling, metabolic, and developmental
pathways in living organisms.  (A small but representative example of
such a network, describing signaling by G protein-coupled receptors,
is shown in Fig. 1.  Other networks under investigation can be found 
online\cite{SethnaGeneDynamics}.)  Tools for the inference of networks from
experimental data, simulation of network behavior, estimation of model
parameters, and quantification of model uncertainties are all
necessary to this endeavor.  Our research into complex biomolecular
networks has revealed an intriguing property, namely their
\emph{sloppiness}: these networks are vastly more sensitive to changes 
along some directions in parameter space than along
others\cite{Brown2003, Brown2004, Waterfall2006, Gutenkunst2006}.
While many tools for the simulation of biomolecular networks have been
built, none support the types of analyses that we need to unravel this
phenomenon of sloppiness.  Therefore, we and our colleagues have
implemented our own software system --
\verb+SloppyCell+ -- to support our research\cite{SloppyCell}.

\verb+SloppyCell+ is an open-source software system written in Python,
which provides support for model construction, simulation, fitting,
and validation.  One important role of Python is to glue together many
diverse modules providing specific functionality.  We use
numpy\cite{NumPy} and scipy\cite{Scipy} for numerics - particularly
for integrating differential equations, optimizing parameters by
least-squares fits to data, and analyzing the Hessian matrix about a
best-fit set of parameters.  We use matplotlib (a.k.a. pylab) for
plotting\cite{Matplotlib}.  A Python interface to the libSBML
library\cite{LibSBML} allows us to read and write models in a
standardized, XML-based file format, the Systems Biology Markup
Language (SBML)\cite{Hucka2003}.  We use the pypar
interface\cite{pypar} to the Message Passing Interface (MPI) library
to coordinate parallel programs on distributed memory clusters.  We
generate descriptions of reaction networks in the \verb+dot+ graph
specification language for visualization via Graphviz\cite{Graphviz}.
We use the \verb+htmllib+ module from the Python standard library to
generate web pages describing models of interest, and the
\verb+smtplib+ module to have simulation runs send email with
information on their status (for those dedicated researchers who
cannot bear to be apart from their work for too long).

While Python serves admirably as a glue layer for a variety of
different tools, as well as a high-level language for specifying and
interrogating simulations, we highlight here in this paper a few
powerful features of Python that enable us to construct highly
dynamic and flexible simulation tools.

\section*{Code synthesis and symbolic manipulation in SloppyCell}

\begin{figure}[t]
\includegraphics[height=3in]{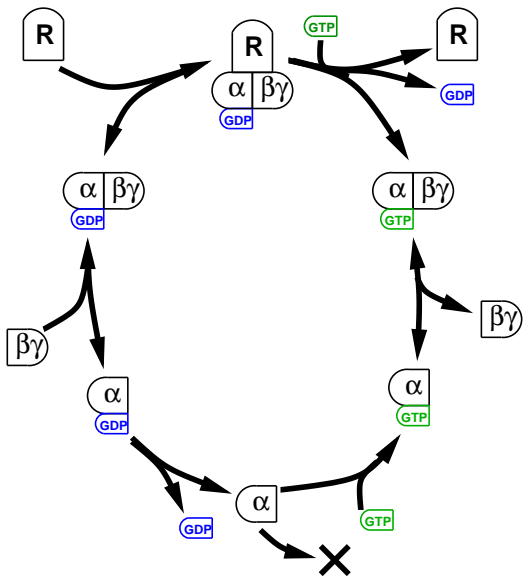}
\caption{\label{GproteinFig}
Model for receptor-driven activation of heterotrimeric G proteins
\cite{Arshavsky2002}.  The $\alpha$ signaling protein is inactive when
bound to GDP and active when bound to GTP.  After forming a complex
with a $\beta\gamma$ protein, binding to the active receptor R allows
the $\alpha$ protein to release its GDP and bind GTP. The complex then
dissociates into back R, $\beta\gamma$, and activated $\alpha$. The
activated $\alpha$ protein goes on to signal down-stream targets,
while the $\beta\gamma$ protein is free to bring new inactive $\alpha$
to the receptor.}
\end{figure}

The dynamics of reaction networks are treated typically as either
continuous and deterministic (modeling the time evolution of molecular
concentrations) or as discrete and stochastic (by simulating many
individual chemical reactions via Monte Carlo).  In the former case,
systems of ordinary differential equations (ODEs) are constructed from
the underlying network topology and the kinetic forms of the
associated chemical reactions.  In practice, these ODEs are often
derived by hand, but they need not be: all the information required
for their synthesis is embedded within a suitably defined network, but
the structure of any particular model is known only at runtime once 
an instance of a \verb+Network+ class has been created and specified.

With Python, we use symbolic expressions (encoded as strings) to
specify the kinetics of different reaction types, and loop over all
the reactions defined in a given network to construct a symbolic
expression for the ODEs describing the time evolution of all chemical
species.  (The reactions are depicted as arrows in Fig. 1; each
reaction can be queried to identify those chemicals involved in that
reaction (represented as shapes), as well as an expression for the
instantaneous rate of the reaction based on model parameters and the
relevant chemical concentrations.) This symbolic expression is
formatted in such a way that we can define a new method,
\verb+get_ddv_dt(y, t)+, that is dynamically attached to an instance
of the \verb+Network+ class using the Python \verb+exec+ statement.
(The term ``dv'' in the method name is shorthand for ``dynamical variables'',
i.e., those chemical species whose time evolution we are solving for.)
This dynamically generated method is then used in conjunction with ODE
integrators (such as \verb+scipy.integrate.odeint+, which is a wrapper
around the venerable LSODA integrator\cite{Hindmarsh1980,
Petzold1983}, or with the variant LSODAR\cite{Hindmarsh1983} that we
have wrapped up within \verb+SloppyCell+ in order to integrate ODEs
with defined events).  We refer to this process of generating the set
of ODEs for a model directly from the network topology as
``compiling'' the network.

This sort of technique allows us to do more than just synthesize 
ODEs for the model itself.  Similar techniques allow us to construct 
\emph{sensitivity equations} for a given model, in order to understand
how model trajectories vary with model parameters.  In order to
accomplish this, we have developed a small package that 
supports the differentiation of symbolic Python math expressions 
with respect to specified variables,
by converting mathematical expressions to
Abstract Syntax Trees using the Python \verb+compiler+ module. 
This allows us to generate another new method,
\verb+get_d2dv_dovdt(y, t)+, that describes the derivatives of the
dynamical variables with respect to both time and the ``optimizable
variables'' (e.g., model parameters whose values we are interested in
fitting to data).  By computing parametric derivatives analytically
rather than via finite differences, we can better navigate the
ill-conditioned terrain of the sloppy models of interest to
us.

The Abstract Syntax Trees (ASTs) created to represent the detailed
mathematical form of biological networks have other benefits as well.
We also use the ASTs to generate \LaTeX\ representations of
the relevant systems of equations.  In practice, this not only saves
us alot of error-prone typing, but it is a very useful tool in
debugging the implementation of a particular model.

Colleagues of ours who are developing \verb+PyDSTool+ -- a Python-based
package for the simulation and analysis of dynamical
systems\cite{PyDSTool} -- have taken this type of approach to code
synthesis for differential equations a step further.  The approach 
described above involves generating Python-encoded right-hand-sides 
to differential equations, which are used in conjunction with compiled 
and wrapped integrators.  For additional performance, \verb+PyDSTool+ 
supports the generation of C-encoded right-hand-sides, which are then 
dynamically compiled and linked using the Python \verb+distutils+ module 
and used with various ODE integrators.

\section*{Parallel programming in SloppyCell}

Because of the sloppy structure of complex biomolecular networks, it
is important not just to simulate a model for one set of parameters,
but over large families of parameter sets consistent with available
experimental data.  We use Monte Carlo sampling to simulate a model
with many different parameter sets in order to estimate model
uncertainties (error bars) associated with predictions.  Parallel
computing on distributed memory clusters efficiently enables these
sorts of extensive parameter explorations.  There are several Python
different packages that provide interfaces to MPI libraries for
message passing, and we have found \verb+pypar+\cite{pypar} to be
especially useful in this regard.

Whereas message passing on distributed memory machines is inherently
somewhat cumbersome and low-level, \verb+pypar+ raises the bar by
exploiting built-in Python support for the pickling of complex
objects.  Message-passing in a low-level programming language such as
Fortran or C typically requires the construction of appropriately
sized memory buffers into which one must pack complex data structures,
but \verb+pypar+ uses Python's ability to serialize (pickle) an
arbitrary object into a Python string, which can then be passed from
one processor to another and unpickled on the other side.  With this
we can pass lists of parameters, model trajectories returned by
integrators, etc.; we also send Python exception objects raised by 
worker nodes back to the master node for further processing.
(These may be generated, for example, if the ODE integrator fails to 
converge for a particular set of parameters.)

Additionally, Python's built-in \verb+eval+ statement makes it easy to
create a very flexible worker that can execute arbitrary expressions
passed as strings by the master (requiring only that the inputs and
return value are picklable). The following snippet of code
demonstrates a basic error-tolerant master/worker parallel computing
environment for arbitrarily complex functions and arguments defined in
some hypothetical module named \verb+our_science+:

\begin{verbatim}
import pypar
# our_science contains the functions we want to execute
import our_science

if pypar.rank() != 0:
    # The workers execute this loop. (The master has rank == 0.)
    while True:
        # Wait for a message from the master.
        message = pypar.receive(source=0)

        # Exit python if sent a SystemExit exception
        if isinstance(message, SystemExit):
            sys.exit()

        # Evaluate the message and send the result back to the master.
        # If an exception was raised send that instead.
        command, msg_locals = message 
        locals().update(msg_locals)
        try:
            result = eval(command)
        except X:
            result = X
        pypar.send(result, 0)

# The code below is only run by the master

# Evaluate our_science.foo(bar) on each worker, getting values for
#     bar from our_science.bar_todo.

command = 'our_science.foo(bar)'
for worker in range(1, pypar.size()):
    args = {'bar': our_science.bar_todo[worker]}
    pypar.send((command, args), worker)

# Collect results from all workers
results = [pypar.receive(worker) for worker in range(1, pypar.size())]

# Check if any of the workers failed, if so, raise the resulting Exception.
for r in results:
    if isinstance(r, Exception):
        raise r

# Shut down all the workers.
for worker in range(1, pypar.size()):
    pypar.send(SystemExit(), worker)
\end{verbatim}

\section*{Summary}

We have very briefly described a few of the fun and flexible features
that Python provides to support the construction of expressive
computational problem solving environments, such as those needed to
tackle complex problems in systems biology.  While any programming
language can be coaxed into doing what is desired with sufficient hard
work, Python encourages researchers to ask questions that they might not
have even thought of were they working in less expressive environments.

\section*{Acknowledgments}

We would like to thank Fergal Casey, Joshua Waterfall, and Robert
Kuczenski for their help in developing and testing \verb+SloppyCell+,
and we would like to acknowledge the insights of Kevin Brown and Colin
Hill in developing predecessor codes which have helped to motivate our
work.  Development of \verb+SloppyCell+ has been supported by NSF
grant DMR-0218475, USDA-ARS project 1907-21000-017-05 , and an NIH
Molecular Biophysics Training grant to RNG (No. T32-GM-08267).

\bibliography{CiSE_python_myers_2}

\end{document}